%% file: Paper1.tex
\documentclass[%
 reprint, amsmath,amssymb, aps,]{revtex4-2}

\usepackage{graphicx}
\usepackage{dcolumn}
\usepackage{bm}
\usepackage{float}
\usepackage{verbatim}
\usepackage{comment}

\begin{document}

\preprint{APS/123-QED}

\title{Microfluidic jet impact: spreading, splashing, soft substrate deformation and injection.}

\author{Diana L. van der Ven$^{1}$}
 \author{Davide Morrone$^{2}$}
\author{Miguel A. Quetzeri-Santiago$^{1}$}
 \email{m.a.quetzerisantiago@utwente.nl}
  \author{David Fernandez Rivas$^{1}$}
  \email{d.fernandezrivas@utwente.nl}
\affiliation{$^1$Mesoscale Chemical Systems group, MESA+ Institute and Faculty of Science and Technology, University of Twente, P.O. Box 217, 7500 AE Enschede, the Netherlands}
\affiliation{$^{2}$Nanovea SRL, Via Balegno 1, 10040 Rivalta di Torino, Italy}

\date{\today}

\begin{abstract}
Injecting with needles causes fear, pain and contamination risks. Billions
of injections every year also cause environmental burden in terms of material consumption and waste. Controlled microfluidic-jet injection systems offer a needle-free alternative. However, understanding the relation between jet parameters and resulting injection depth are needed to enable targeting specific skin layers, and enhance the pharmacokinetics of various therapeutic compounds. The complexity of skin, its opacity and non-linear mechanical properties, pose a technological challenge. Hence the use of surrogates is instrumental to understand how to inject without needles. In particular, reducing undesired splashing upon jet impact and liquid squeeze-out after injection are needed to minimize infection risks and ensure accurate dosage. Therefore, in this paper we explore how microfluidic jet characteristics influence the impact outcome on a range of materials as skin surrogate. Jets with velocities between 7 - 77 m/s and  diameters 35 - 130 $\mu$m were directed at substrates with shear moduli between 0.2 kPa and 26 GPa. We found seven different regimes depending on jet inertia and substrate shear modulus. Furthermore, three distinct transition regions were identified as the thresholds between regimes: i) spreading/splashing threshold, ii) dimple formation threshold, and iii) plastic/elastic deformation threshold. These thresholds allow predicting the required jet velocity and diameter to inject substrates with known shear modulus. We found that jet velocity is a better predictor for the injection depth compared to the Weber number, as the jet diameter does not influence the injection depth. Our findings are relevant for advancing needle-free injection research, because the shear modulus of skin depends on multiple factors, such as ethnicity, body part and environmental conditions.
\end{abstract}

\maketitle
\section{Introduction}\label{sec:level1}
Solid needle injections have inherent drawbacks despite their wide use and effectiveness in medical and cosmetic procedures. Worldwide, 16 billion needles are disposed yearly\cite{ref1}, with an additional 12 billion due to the COVID-19 vaccinations\cite{ref2}, creating an environmental burden in terms of material usage and waste\cite{refIE1, refIE2}. Needles pose infection risks upon sticking incidents\cite{ref3} and re-use, the latter being estimated to cause over 500 million deaths yearly\cite{ref4}. Additionally, injections cause pain, and fear is experienced by 22$\%$ of the adult population\cite{ref5}. This fear may result in avoidance of treatment\cite{ref6,ref7}, decreasing the effectiveness of vaccination programs\cite{ref8}.
To address these issues, alternatives to conventional injections are being developed, such as liquid jet injections. Liquid jets can be made by accelerating fluids with various mechanisms\cite{ref9,ref10,ref11}. Most commercialized injectors can only perform injections at a minimum depth of $\approx$~1~mm, and are associated with pain and bruising, limiting their appeal in replacing needles\cite{ref12,ref13}. Controlled jet injections into shallow depths in the epidermis may result in reduced tissue damage, causing less bruising and pain compared to the current alternatives\cite{ref14,ref15,ref16,ref17}. Moreover, control over injection depth could improve the pharmacokinetics of various therapeutics. For example, vaccines delivered within the epidermis require five to ten times less dosing due to the abundance of specialized immune cells\cite{ref18,ref19}. Also, dermal insulin injections are reported to be more efficient compared to traditional subcutaneous delivery\cite{ref20}.
The relation between jet characteristics and injection depth in the skin must be understood to enable controlled injections. Correlating input parameters with the injection depth can be challenging as skin is a highly complex, multilayered tissue\cite{ref21}. The mechanical response of skin is dependent on measurement type, leading to reported skin stiffness values ranging from 1~MPa – 1~GPa\cite{ref22,ref23}. Additionally, skin’s poro-viscoelastic behavior causes a portion of the injected liquid to be ejected during, or after, injection, due to the elastic recovery of the material (squeeze-out)\cite{ref24,ref25,ref26,ref27}. Multiple papers show how jet parameters relate to injection depth\cite{ref13,ref24,ref28,ref29}, dispersion\cite{ref24,ref29,ref30,ref31}, and delivery efficiency\cite{ref15,ref24,ref29}. However, a detailed study of  the impact behavior, especially the breakup of a jet upon impact (splashing) and squeeze-out at the microscale is lacking.
Splashing and squeeze-out must be minimized as both reduce the delivery efficiency, and pose infection risks in extreme cases\cite{ref33,ref34}. Minimizing splashing requires knowledge about microjet impact behavior and how this relates to jet break-up. Droplet splashing has been extensively characterized\cite{ref35,ref36,ref37,ref38,ref39,ref40,ref41,ref42,ref43,ref44}, and it has been shown to depend on the substrate, liquid characteristics, and ambient pressure\cite{ref35,ref36,ref45,ref46}. Furthermore, it is known that microscale droplets show different impact behavior compared to macroscale droplets as the rim of liquid formed after the impact becomes comparable to the mean free path of air molecules \cite{ref47,ref48,ref49}.
In contrast to droplet splashing, studies on jet impact and splashing are limited, and reported values for the splashing threshold vary widely\cite{ref50,ref51}. Furthermore, few studies have explored the splashing dynamics for substrates with different storage moduli\cite{ref52,ref26,ref53}, and it remains unknown how microscale jet behavior compares to macroscale jets. Therefore, in this paper, we study microfluidic-jet impact and injection behavior onto a range of substrate stiffness to gain fundamental knowledge on fluid dynamics and soft matter response. This knowledge can assist in optimizing controlled needle-free injections by controlling injection depth and reducing splashing and squeeze-out.

\section{\label{sec:level2}Materials and Methodology}
\subsection{\label{sec:level2.1}Methodology}
A continuous wave (CW) laser setup was employed to perform controlled injections with microfluidic jets using thermocavitation. A vapor bubble is created upon CW laser beam exposure at the glass-liquid interface due to the conversion of laser energy to heat\cite{ref54,ref55}. The bubble expands and pushes the liquid out of the chip as a liquid jet. By changing the channel design, the channel filling level, the input laser power, laser pulse duration and distance from the laser focal point to the base of the chip, jets with varying characteristics can be obtained\cite{ref16,ref56}. The used chips, some of their details, and the resulting jet ranges are shown in Table 1. Figure 1 shows an illustration of a microfluidic chip and a jetting event.

\begin{figure*} 
\includegraphics[width=0.75\textwidth]{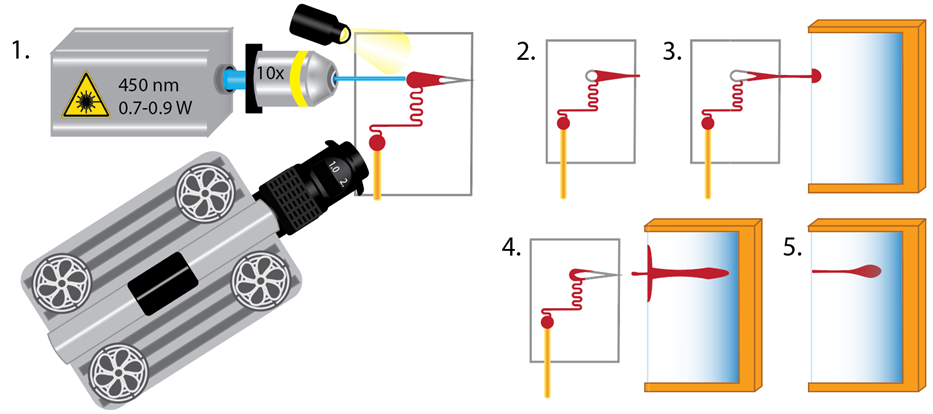}
\caption{The laser is focused on the base of the microfluidic chip chamber (1). Upon activation, the laser heats up the liquid inside the chamber, generating an expanding bubble (2). The expanding bubble pushes the liquid in front of it and generates a microfluidic jet (2 - 3). The jet impacts the surface of the agarose gel (3), rupturing it and injecting if the impact velocity is larger than the threshold value (4). Over time, the formed cavity contracts and the injectate compacts (5). Here, the setup is depicted for shooting angular movies, to visualize the surface. When viewing the substrate from the side, the camera is perpendicular to the chip.}
\label{fig:wide}
\end{figure*}
\begin{table*}[!htp]    
\caption{\label{tab:table1}%
An overview of the dimensions and characteristics of the used chips (channel length, used channel filling levels, channel taper angel, orifice width, channel wall shape) and corresponding jet characteristics: diameter ($D_{jet}$), velocity ($V_{jet}$) and Weber number ($We_{jet}$).}
\begin{ruledtabular}
\begin{tabular}{ccccccccc}Chip&Channel length&Filling levels&Taper angle&Orifice width&Channel wall&$D_{jet}$&$V_{jet}$&$We_{jet}$\\
\hline
\ 1 & \textrm{1.9 mm} & \textrm{0.4-1.9 mm} & 7.5 & 50 $\mu$m & Flat&45-130 $\mu$m& 7-37 m/s& 70-2500 \\
\ 2 & \textrm{1.9 mm} & \textrm{0.9-1.9 mm} & 7.5 & 50 $\mu$m & Flat&40-110 $\mu$m& 17-67 m/s&1200-4000 \\
\ 3 & \textrm{2.1 mm} & \textrm{0.7-1.9 mm} & x & 500 $\mu$m &Curved&45-100 $\mu$m& 71-77 m/s&1600-4000 \\
\end{tabular}
\end{ruledtabular}
\end{table*}

\subsection{\label{sec:level2.2}Experimental setup}
The experimental setup consists of a glass microfluidic chip filled with a water-dye solution and a CW laser diode ($\lambda$ = 450 nm) focused on the glass-liquid interface using a 10x objective (Olympus) (see figure 1). An Arduino controller sets the laser pulse duration (10-30~ms)\cite{ref16}. The chips were designed and fabricated from MEMPax Borofloat glass wafers (Schott) under cleanroom conditions\cite{ref56}. The chips are aligned with the test substrates at a stand-off distance $X_{s}$~=~2~mm. A red dye (Direct Red 81, Sigma, CAS: 2610-119) was dissolved at 0.5 wt$\%$ in deionized water (${d}H_{2}O$) to maximize the laser energy absorption of the liquid. The solution had a density, $\rho$ = 1000 $kg/m^{3}$; viscosity, $\eta$~=~0.91~mPaS; and surface tension, $\sigma$ = 47 mN/m, as determined at 22°C and described elsewhere\cite{ref16}. 

 Nucleation, jetting, impact and injection events are imaged using a high-speed camera (Photron FASTCAM SA-X2) with a mounted Navitar (12x Ultra Zoom) at 2x magnification captured at 1.8×10$^{5}$ or 1.2×10$^{5}$ frames per second (fps), and backlight illuminated using a SCHOTT light source (CV-LS series) with flexible light guide. The field of view was either 640x128 or 768×64 pixels with a pixel size of 10 $\mu$m. A mounted long pass 495 nm filter (Thorlabs, FGL495M) was used to avoid over saturation from the laser light. 

\subsection{\label{sec:level2.3}Jet impact substrates}
\subsubsection{Glass slides}
Glass slides (MEMPax Borofloat, 30×15×0.5mm) were used to test the impact on stiff substrates. The slide was attached to a XYZ-stage at level surface. The glass surface was cleaned with acetone before and after jet impact using a Kim-wipe. The liquid contact angle was determined by placing a 5 $\mu$L droplet on the glass surface and captured with the Photron camera. The contact angle was calculated by taking half the reflection angle (processed using ImageJ) which resulted in 23° for water and 15° for the dye solution. The glass Young’s modulus (E) is 62.6 GPa, with a Poisson’s ratio ($\nu$) of 0.196\cite{ref57}. As substrate E was known, the Poisson’s ratio was used to calculate G, using: 
\begin{eqnarray}
E=2G(1+\nu)
\end{eqnarray}
Where E is the elastic modulus, G is the shear modulus and v is the Poisson’s ratio. This resulted in a shear modulus (G) of 26 GPa (see Table 2).

\subsubsection{Agarose gel preparation and casting}
Agarose (Sigma, CAS: 9012-36-6) was weighed on a microbalance (VWR) and mixed with ${d}H_{2}O$. Five concentrations were prepared (0.125, 0.25, 0.5, 1 and 2 wt$\%$) and the flask containing the solution was closed using a Precision seal rubber septa (Merck) to minimize evaporation, and was pierced with an injection needle to prevent pressure build-up. The agarose was dissolved using a magnetic stirrer/heater plate (IKA-CMAG), stirred and gently boiled until the solution was completely transparent. The solution was pipetted into a 3D printed mold (3.4×1.8×0.9 cm, $\approx$1.7~mL) until a convex gel surface was obtained. The mold has glass slides on either side to create visual access. After cooling down the solution solidified into a transparent and homogeneous gel which were cut with a surgical grade scalpel to create a flat gel surface.

\subsubsection{Agarose gel characterization}
Nanoindentation tests were performed on 0.5, 1 and 2 wt$\%$ agarose using a Nanovea PB1000 Mechanical tester equipped with a 400 mN Nano module (Nanovea, Irvine, CA, USA), actuated with a piezo driver and equipped with direct load and depth sensors. The high-resolution Nano module allows the application of low target loads (100-200 $\mu$N) using a 200 $\mu$m radius 90° sphero-conical indenter to evaluate elastic moduli of the samples. The system monitors the applied load with an independent load-cell and records the indenter position within the material using an independent capacitive depth sensor. A high-speed multichannel 24-bit acquisition card drives both the actuator and the sensors. E was determined to be 195.09 ± 23.29, 43.31 ± 1.42, 10.57~±~0.79~kPa for 2$\%$, 1$\%$ and 0.5$\%$ respectively. As 0.25$\%$ and 0.125$\%$ were too soft to obtain a reliable measurement, E was calculated based on the trend of 0.5-2$\%$ ($y=44.352x^{2.103}, R^{2}~=~0.9997$), giving 2.4 and 0.6~kPa. See Appendix A for more information. With equation (1), for the listed E, using v= 0.5\cite{ref58}, G was calculated. All values are listed in Table 2.

\begin{table}[H]
\caption{\label{tab:table2}%
Overview of the used test substrates (Borosilicate glass, and 2$\%$, 1$\%$, 0.5$\%$, 0.25$\%$, 0.125$\%$ agarose) and relevant characteristics: the Young’s modulus and the Poisson’s ratio that was used to calculate the shear modulus.}
\begin{ruledtabular}
\begin{tabular}{lccc}Substrate&E&$\upsilon$&G\\
\hline
Glass & 62.6 GPa & 0.196 & 2.6 GPa\\
2$\%$ agarose & 195.1 kPa & 0.5 & 65 kPa\\
1$\%$ agarose & 43.3 kPa & 0.5 & 14 kPa\\
0.5$\%$ agarose & 10.6 kPa & 0.5 & 3.5 kPa\\
0.25$\%$ agarose & 2.4 kPa & 0.5 & 0.8 kPa\\
0.125$\%$ agarose & 0.6 kPa & 0.5 & 0.2 kPa\\
\end{tabular}
\end{ruledtabular}
\end{table}

\section{\label{sec:level3}Results and discussion}
The large range of jets showed distinct impacting behavior depending on jet and substrate characteristics. We observed seven regimes, and characterized them in terms of the Weber number of the impacting jet ($We_{jet}$) and substrate G. As the $We_{jet}$ describes the relative contribution of the fluid’s inertia compared to the surface tension, it is often used to determine splashing thresholds of impacting liquids\cite{ref35,ref36,ref37,ref50}.\\
For these studies $We_{jet}$ is defined as: 
\begin{eqnarray}
We_{jet}=\frac{(\rho  V_{jet}^2  D_{jet})}{\sigma}
\end{eqnarray}
where $\rho$ is the fluid density, $V_{jet}$ is the jet velocity, $D_{jet}$ is the jet diameter, and $\sigma$ is the surface tension. In these experiments, both $\rho$ and $\sigma$ are constant, so $We_{jet}$ is a function of $V_{jet}$ and $D_{jet}$. 

First, the seven regimes will be reviewed in detail, using illustrations and snapshots. Movies of all described regimes can be found in the online supplemental information. Next, all regimes are combined into a regime map, and the the relation between $We_{jet}$ and substrate G is discussed. Finally, we review how $We_{jet}$ relates to injection depth.

\subsubsection*{Regime 1: Jet spreading}
For impact on glass (G = 26 GPa, $We_{jet}<1150$) and 2$\%$ agarose (G = 65 kPa, $We_{jet}<$1290), we did not observe material deformation or jet splashing (Figure 2 and Figure 3, top). Impact duration depends on ejected volume, $D_{jet}$ and $V_{jet}$, and lasts for 500 $\mu$s on average. The jet spreads on the surface upon impact (Figure~2, t~=~6~$\mu$s), creating a smooth film that spreads radially over the surface and flows with wave formation (t~=~11-22 $\mu$s). This is comparable to the deposition behavior described for droplets and macroscale jets impacting solids\cite{ref50,ref59}. As impact progresses (t~=~83~$\mu$s), the amplitude of the waves increases and lamella formation is seen (Figure 2, red arrow), sometimes to the point where asymmetric fingering edge formation can be seen, but no secondary droplets detach from the expanding liquid sheet (Figure 2, red circles). This fingering behavior shows similarities to the fingering and subsequent droplet break-up reported for skating thin films on solid surfaces\cite{ref60}. At the end of impact (t~=~367 $\mu$s) the film flow shows a decrease in frequency and amplitude of waves.

\begin{figure}
\includegraphics[width=\columnwidth,height=\textheight,keepaspectratio]{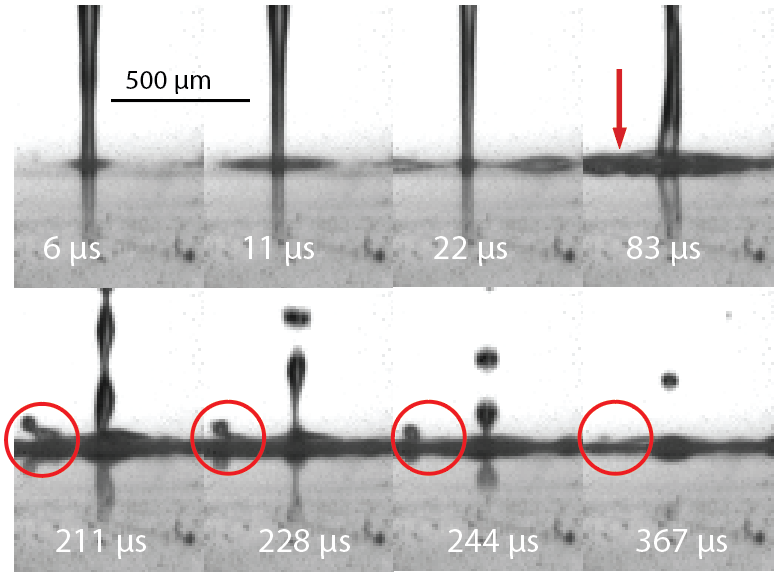}
\caption{\label{fig2} Jet spreading. Upon impact (t = 6 $\mu$s) the jet spread without splashing. During impact, a lamella formation is seen (t = 83 $\mu$s, red arrow), followed by liquid surface instabilities (red circle, t = 211-467) for higher $We_{jet}$, yet these instabilities do not result in splashing, as no droplet release is seen. $We_{jet}$~=~960 ($V_{jet}$ =30 nm/s,$D_{jet}$ = 50 $\mu$m). Please note that the jets do not penetrate the surface, instead the refection of the jet on the glass is seen. See Supplemental Movie 1 to view the full event.}
\end{figure}
\begin{figure}[ht!]
\includegraphics[width=\columnwidth,height=\textheight,keepaspectratio]{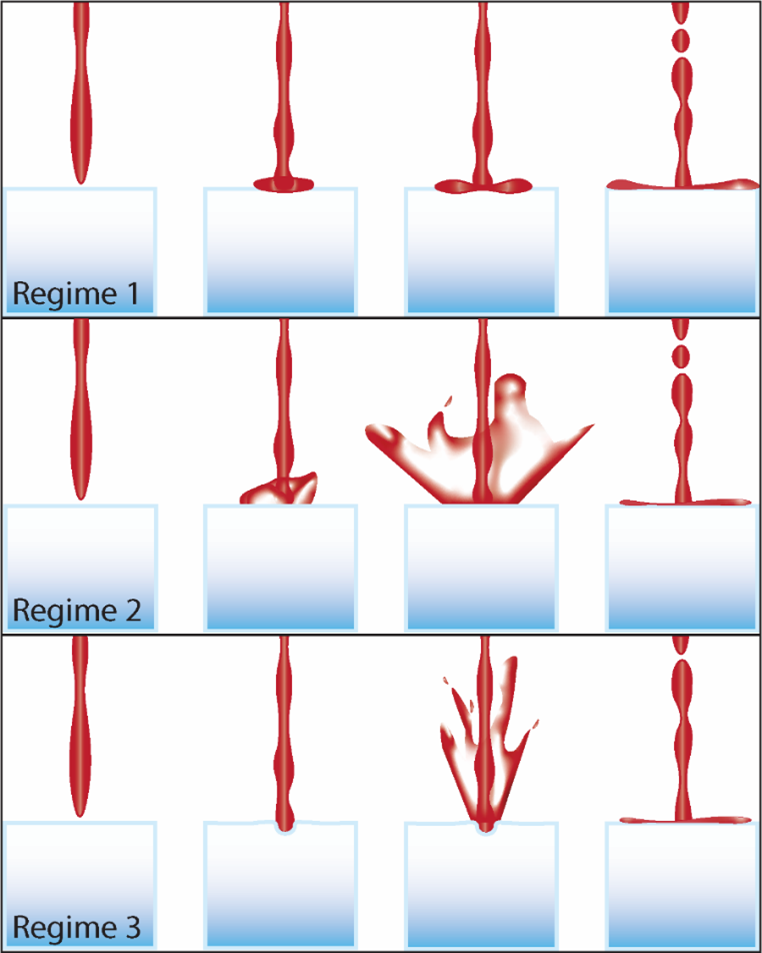}
\caption{\label{fig3} Diagrams showing elastic substrate deformation regimes. 
Regime 1) Jet spreading: the jet impacts the substrate without visible deformation. The jet spreads and forms a pool without splashing. This Regime is observed for G~=~26~GPa, $We_{jet}$ $<$1150 and G = 65 kPa, $We_{jet}<$1290. Regime 2) Jet splashing: the liquid jet impacts the surface and is not deformed, the jet splashes. Observed  for: G~=~26~GPa, $We_{jet}>$1340 and G = 65 kPa, $We_{jet}>$1390. Regime 3) Surface defeat: the jet impacts, creates a dimple in the surface and splashes at an angle corresponding to the dimple geometry. Observed for: G = 65 kPa, $We_{jet}>$3335; G~=~14~kPa, $We_{jet}$ = 400-1970; G = 3.5 kPa, $We_{jet}<$1060 and G~=~0.8~kPa, $We_{jet}<$170. Supplemental Movies 1-3 show examples of the corresponding Regimes.}
\end{figure}

\subsubsection*{Regime 2: Jet splashing}
For impact on glass (G = 26 GPa, $We_{jet}>$1340) and 2$\%$ agarose (G = 65 kPa, $We_{jet}>$1390) splashing was observed in the absence of visible dimple formation, as illustrated in Figure 3. Almost directly upon impact (Figure 4, top, t = 11  $\mu$s), outward splashing is seen (150-750~$\mu$m). At $We_{jet}$ $<$ 3500, liquid-film expansion occurs similarly to what has been described for Regime~1 (Figure 4, bottom),  whereas at $We_{jet}$ $>$ 3500 fingering (Figure 4, top, t = 89 $\mu$s) and subsequent droplet release is seen (t = 139 $\mu$s). This indicates that for $We_{jet}$ $>$ 3500 the jet-tail has sufficient inertia to create secondary splashing within the liquid pool formed by the jet after impact, while jets impacting at $We_{jet}$ $<$ 3500 lack inertia, causing only outward splashing upon impact. In accordance with the findings of Howland et al., outward splashing occurred at lower $We_{jet}$ for impact on glass as compared to agarose \cite{ref37}.
Splashing, the lamella formation and subsequent fingering can be described as follows. As the spreading front of the liquid rim advances, a lubrication force from the surrounding gas pushes the lamella upwards. If the lubrication force is larger than the surface tension the lamella will break up into secondary droplets\cite{ref61}. Therefore the mean free path of the gas molecules ($\lambda$), the lamella thickness and speed determine the splashing phenomena. 

If the rim diameter of the lamella increases faster than the lamella lifts from the substrate, the droplet will not splash\cite{ref49}. Likewise, for microdroplet impacts the lamella thickness is in the order of magnitude of $\lambda$, making the airlift force negligible\cite{ref47}. The threshold for inhibiting splashing was reported to be $We_\lambda=~\rho~V^{2}~/ \sigma~\geq~0.5$, where $\lambda=6.9 \times 10^{-5}$ m is the mean free path of the air at atmospheric pressure\cite{ref47}.
\begin{figure}[ht!]
\includegraphics[width=\columnwidth,height=\textheight,keepaspectratio]{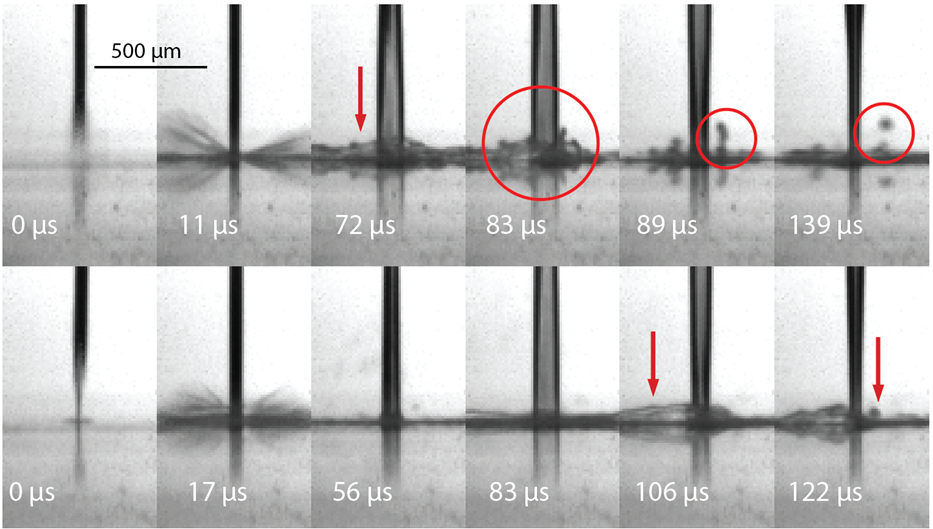}
\caption{\label{fig4} Top) Jet double splashing upon glass. The jet splashes upon impact (t = 11 $\mu$s), and during impact a lamella forms (72 $\mu$s, red arrow), which starts to break-up (t =~83~$\mu$s, big red circle) and shows fingering (t= 89 $\mu$s, red arrow) resulting into secondary droplet release (139 $\mu$s, red circle). $We_{jet}$ = 290 ($V_{jet}$ = 16 m/s,$D_{jet}$ = 55  $\mu$m). Supplemental Movie 4.1 shows the full event. Bottom) Jet single plashing. The jet splashes upon impact (t = 17 $\mu$s). During further impact lamella formation is seen (t = 106 $\mu$s, red arrow), which may cause some instabilities (t = 122 $\mu$s, red arrow) but does not result in secondary droplet release. $We_{jet}$ = 130 ($V_{jet}$~=~11m/s, $D_{jet}$ = 50 $\mu$m). See Supplemental Movie 4.2}
\end{figure}

Qian et al. reported a jet splashing regime at $We_{jet}$ = 617 ($D_{jet}$~=~6~mm)\cite{ref50}, almost twice as low compared to our findings ($We_{jet}$~=~1340). In contrast, Trainer et al. saw splashing of jets ($D_{jet}$ = 3 mm) above $We_{jet}$= 1500\cite{ref62}, closer to our findings. Therefore, we interpret that the low splashing threshold found by Qian et al. is due to the disturbances observed in the jets that lead to break-up\cite{ref50}.

\subsubsection*{Regime 3: Surface defeat}
This Regime is seen for impact on 2$\%$ agarose (G~=~65~kPa, $We_{jet}>$3335), 1$\%$ agarose (G~=~14~kPa, $We_{jet}$ = 400-1970), 0.5$\%$ agarose (G = 3.5 kPa, $We_{jet}<$1060) and 0.25$\%$ agarose (G = 0.8 kPa, $We_{jet}<$170) and illustrated in Figure 3. The jet splashes upon impact (t~=~0 $\mu$s), as seen in Regime 2, and subsequently creates a dimple in the substrate, followed by a secondary splash reaching heights $>$ 750 $\mu$m. Uth et al. reported a similar situation for macro-jet ($D_{jet}$ = 2 mm) impact on ceramics and gels\cite{ref63}, where the secondary splash is caused by backflow. Indeed, upon impact, the jet creates a dimple with a depth $\approx$ 0.5 $D_{jet}$, and flows back as it lacks force to further deform the surface. Additionally, the dimple geometry defines the backflow angle, as reported earlier\cite{ref63}, However, due to lack of resolution it is out of scope for these studies to give an exact relation. Finally, as impact progresses (t $\approx$ 400 $\mu$s), the dimple shrinks and disappears, while the final part of the jet impacts without splashing or injection.

\subsubsection*{Regime 4: Splash, injection and squeeze-out}
This Regime was found for 1$\%$ agarose (G~=~14~kPa, $We_{jet}<$ 2145), 0.5$\%$ agarose (G = 3.5 kPa, $We_{jet}$= 1240-3050) and 0.25$\%$ agarose (G = 14 kPa, $We_{jet}<$2145) and is depicted in Figure 5. First, an initial dimple forms, followed by splashing that reaches heights of $\approx$ 750 $\mu$m), with an angle determined by the geometry of the dimple. As jet impact continues (t $\approx$ 200 $\mu$s), the dimple deepens and the substrate surface yields, causing the jet to penetrate further into the substrate while the splashing ceases. Uth et al. also found a decrease in splashing along the injection duration\cite{ref63}.

Towards the end of the jet impact the injection depth no longer increases (t $\approx$ 400 $\mu$s). Liquid jets have a non-uniform velocity throughout the jet due to air resistance. This causes the jet tail to have a lower speed compared to its front, and consequently the impact pressure of the jet decreases over time. Moreover, agarose is a viscoelastic material\cite{ref64}, and its viscous response suppresses the elastic component temporarily during impact, allowing jet penetration. Once the elastic recovery occurs, as the jet impact force stagnates towards the end of the injection (t $\approx$ 400 $\mu$s), the gel returns to its original position, thereby squeezing out the liquid that penetrated the gel\cite{ref65}.

\begin{figure}
\includegraphics{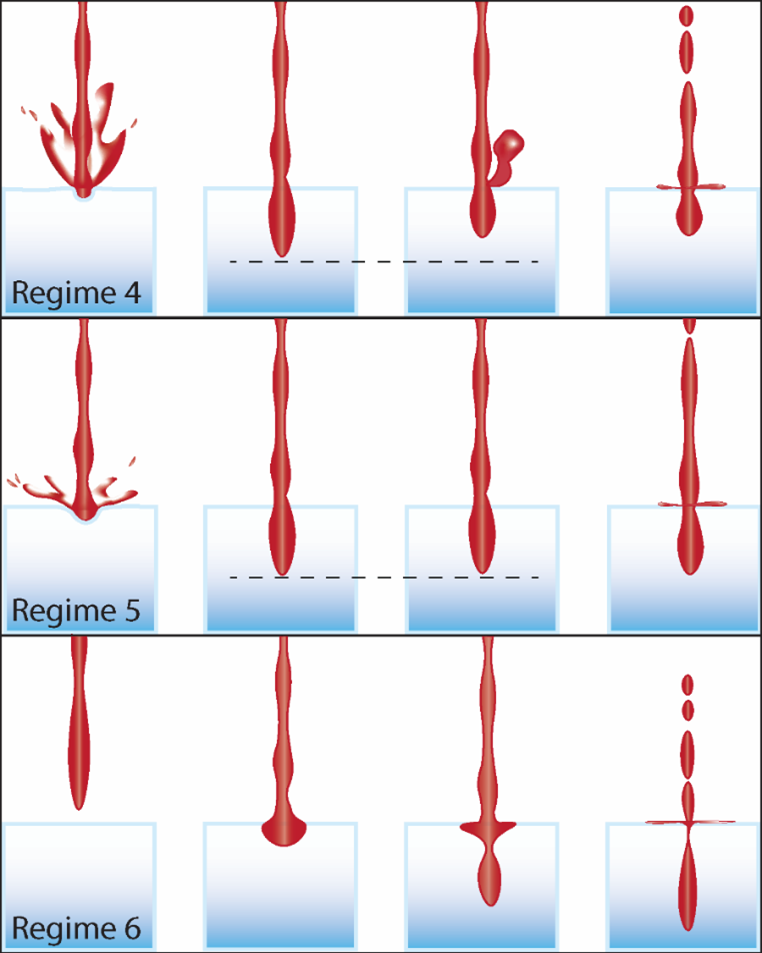}
\caption{\label{fig5} Diagrams showing plastic substrate deformation regimes. Regime 4) Splash, injection and Squeeze-out: The jet impact on the agarose gel surface creates a dimple and eventually penetrates the surface. Before penetrating the jet splashes similarly to the surface defeat regime. After the injection, part of the injected liquid is expelled because of the viscoelastic properties of the gel. This Regime is observed for: G = 14 kPa, $We_{jet}<$ 2145; G = 3.5 kPa, $We_{jet}$ = 1240-3050) and G = 14 kPa, $We_{jet}<$ 2145. Regime 5) Splash and injection: This regime resembles Regime 4 in the first stages; there is jet splashing upon impact, and a dimple is formed and penetrated, but no liquid is expelled at the end of the process. This Regime is observed for: G = 35 kPa, $We_{jet}>$~3375 and G = 0.8 kPa, $We_{jet}$ = 655-1240. Regime 6) Clean injection: The jet impacts, deforms the surface and penetrates it. In this Regime the jet does not splash. This is the ideal situation for needle-free jet injection systems. This Regime is observed for: G = 0.8 kPa, $We_{jet}>$ 3375 and G~=~0.2~kPa, $We_{jet}<$ 270. Supplemental Movies 4-6 show examples of the corresponding Regimes.}
\end{figure}

\subsubsection*{Regime 5: Splash and injection}
This Regime, as depicted in Figure 5, was found for impact on 0.5$\%$ agarose (G = 35 kPa, $We_{jet}>$~3375) and 0.25$\%$ agarose (G = 0.8 kPa, $We_{jet}$ = 655-1240). A dimple is formed with minor splashing (500 – 100 $\mu$m), followed by substrate penetration. For agarose concentrations lower than 0.25$\%$ injection is seen at lower velocities ($V_{jet}>$30 m/s) as compared to the agarose concentrations ($>$ 0.25$\%$) $V_{jet}>$ 55 m/s. Since intermolecular forces are larger for higher agarose concentrations\cite{ref58,ref64,ref65}, jet impact deforms the substrate instead of producing a splashing sheet. Towards the end of impact (t~$\approx$~400~$\mu$s), the injected area may compress or recede, but no squeeze-out is seen in contrast to Regime 4. 

Thus, in this Regime we expect to reach the plastic deformation of the material, i.e., jet inertia overcomes the elasticity of the material beyond the recovery point. Additionally, while the jet penetrates (t $\approx$ 400-600 $\mu$s), the dimple in the surface flattens and widens, eventually pinching off from the injection cavity, sealing the injectate within the gel. 

We can compare our results by looking at studies where high-speed projectiles were impacted on various concentrations of gelatin gels, and cavity regimes were defined based on projectile elastic Froude number ($Fr$) as function of substrate G\cite{ref66}. Within the reported cavity types our observations can be best compared to the shallow seal regime, which we found for G = 3.5 kPa and $We_{jet}>$ 3500, $Fr_{e}>$ 1200; G~=~0.8~kPa and $We_{jet}>$ 650, $Fr_{e}>$ 1000; G = 0.2 kPa and $We_{jet}$~=~125-215,  $Fr_{e}>$ 700-1300. Our results are in agreement with Kiyama et al., where they found shallow seals starting from $Fr_{e}$~ 200 – 400 and transitioning to surface seal at $Fr_{e}>10^{4}$ \cite{ref66}. We did not observe a transition to a surface seal as for our experiments $Fr_{e} < 10^4$.

\subsubsection*{Regime 6: Clean injection}
This Regime is the optimal scenario for injections and it was observed for impacts at 0.25$\%$ agarose (G = 0.8 kPa, $We_{jet}>$ 3375) and 0.125$\%$ agarose (G = 0.2 kPa, $We_{jet}<$ 270). Like Regime 5, no splashing is seen because the jet-impact pressure only deforms the substrate (see Figure 5). Compared to Regime 5, the dimple expands faster and wider (200 $\mu$s, 20-40 times $D_{jet}$ for Regime 6 versus 400 $\mu$s, 1.5-2 times $D_{jet}$ for Regime 5).

Observing at an angle, different phases in the cavity dynamics can be highlighted (see Figure 6). First, the jet front forms a shallow dimple (t = 8 $\mu$s) which expands into a bell-shaped cavity at t = 25-125 $\mu$s. The formation of cavities wider than $D_{jet}$ is caused by pressure build-up of the impacting jet at the interface via momentum transfer from the jet, causing the pressure to move radially outwards and entraining the surrounding air\cite{ref66}. Subsequently (t~=~258~$\mu$s), the trailing part of the jet penetrates the gel surface forming a narrow secondary cavity, which finally seals the jet within the gel (t = 633 $\mu$s). For 0.25$\%$ agarose, the cavity-diameter is approxiamtely 20 times $D_{jet}$, whereas in 0.125$\%$ the cavity-diameter is around 40 times $D_{jet}$ (t = 258 $\mu$s).
\begin{figure}
\includegraphics[width=\columnwidth,height=\textheight,keepaspectratio]{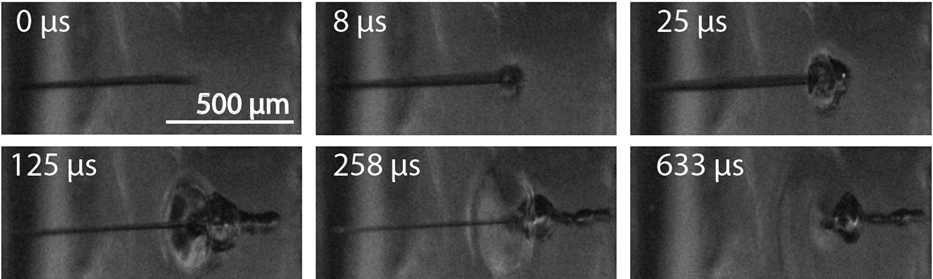}
\caption{\label{fig6} Snapshot sequences of a jet ($We_{jet}\approx$1000) impacting on a 0.2 kPa substrate (0.125$\%$ agarose) in the clean injection regime. Upon impact (t = 0 $\mu$s) the jet deforms the agarose gel (t= 8 $\mu$s) and  penetrates the surface (t = 25 $\mu$s) causing a wide cavity at the surface (t = 125-258 $\mu$m). At the end of the impact (t = 633 $\mu$s) the cavity seals and the liquid remains in the gel. In this sequence the $V_{jet}\approx$  30 m/s, $D_{jet}$ = 50 $\mu$m and the substrate is 0.125$\%$ agarose.}
\end{figure}

\subsubsection*{Regime 7: Splashing substrate (no jet splashing)}
This Regime corresponds to the softest substrate, 0.125$\%$ agarose (G = 0.2 kPa, $We_{jet}<$ 290), see Figure~7. In this regime, the substrate splashes upon jet-impact, similarly to water entry experiments, i.e., secondary droplets detach from the
substrate, while no jet-splashing is seen. For water entry experiments this regime has been observed at $We$~$\approx$~180-230\cite{ref56,ref57}.

\begin{figure}[h]
\includegraphics[width=\columnwidth,height=\textheight,keepaspectratio]{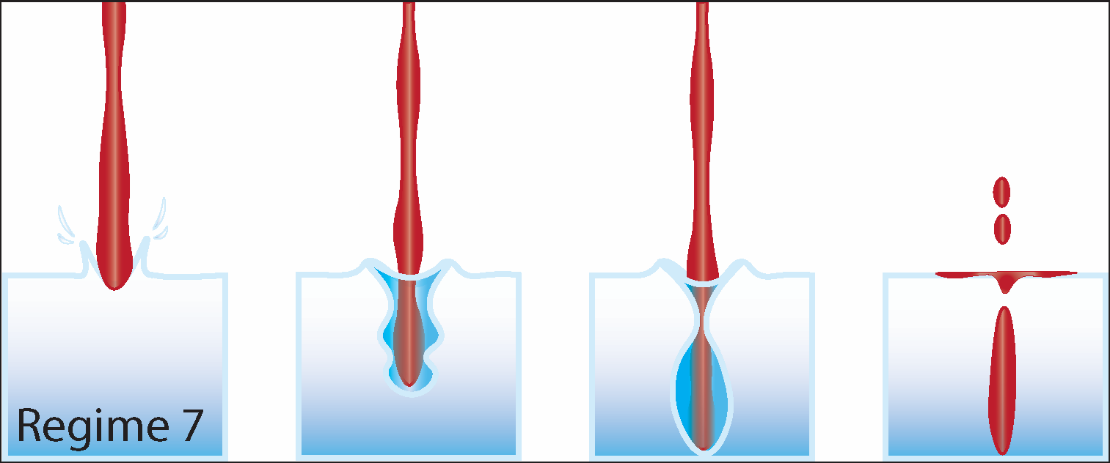}
\caption{\label{fig7} Regime 7) Splashing Substrate: In this Regime, the jet does not splash upon impact, but the substrate does. This Regime was found only for: G = 0.2 kPa, $We_{jet}<$ 290. Supplemental Movies 7.1-7.2 show examples of the described Regime.}
\end{figure}

As the jet starts penetrating the gel, the gel surface bulges (Figure 8, t = 6 $\mu$s), and we observe a crown formation of the agarose sheet from where droplets detach (t~=~11-17 $\mu$s). This crown formation resembles water entry experiments. However, higher $We$ is required to obtain crown formation in agarose gels as compared to water. This was expected as viscous dissipation is larger in agarose gels than in water, and viscous dissipation stabilizes the crown rim\cite{ref38,ref67}. Furthermore, agarose has a larger effective surface tension than water, and may contribute to the higher crown formation threshold\cite{ref67}.
As jet-impact progresses, the substrate-splashing ceases while the bulge remains and widens (t = 28-39 $\mu$s). At the end of the impact (t = 900 $\mu$s), the surface recovers to its original position. The bulging substrate-surface results from pressure build-up and momentum transfer from the impacting jet. Besides the substrate splashing upon jet-impact and the bulging surface, the injection and cavity dynamics are very similar to Regime 6.

\begin{figure}
\includegraphics[width=\columnwidth,height=\textheight,keepaspectratio]{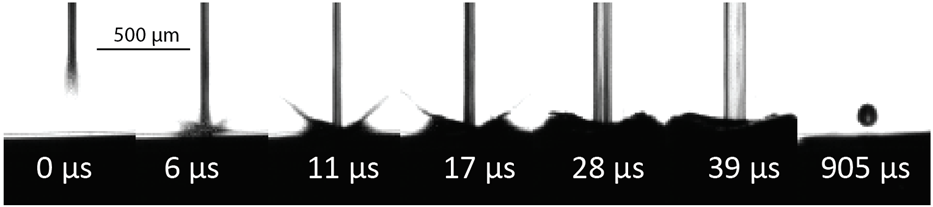}
\caption{\label{fig8} Snapshots of the splashing substrate (Regime~7). At G =~0.2~kPa (0.125$\%$ agarose) and $We_{jet}$=2600 ($V_{jet}$~=~55m/s, $D_{jet}$ = 40 $\mu$m). As the jet starts penetrating the gel-surface, the surface bulges (t = 6 $\mu$s) creating a crown from where secondary droplets detach (11 - 17 $\mu$s). As jet-impact progresses, the substrate-splashing ceases while the bulge remains and widens (t = 28 - 39 $\mu$s). Upon the end of the impact (t = 900 $\mu$s), the surface recovers to its original position. Supplemental Movie~7.1 highlights the cavity dynamics, and 7.2 corresponds to this Figure.}
\end{figure}

\subsubsection*{Regime map}
In Figure 9 we show a regime map by plotting $We_{jet}$, in terms of the substrate shear modulus on a log-log scale. The blurred symbols represent the experimental data, and the larger symbols show the first transition value to a new regime. Both the transition from spreading to splashing and the transitions to different regimes show linear trends and we have derived three thresholds, defined as: spreading/splashing threshold: $We_{jet}=1645 G^{-0.022}(y_{1})$; dimple formation threshold: $We_{jet}=0.017 G^{1.1} (y_{2})$; and plastic/elastic deformation threshold: $We_{jet}=0.728 G^{0.83} (y_{3})$.

\begin{figure*}[!htp]
\includegraphics[width=12cm]{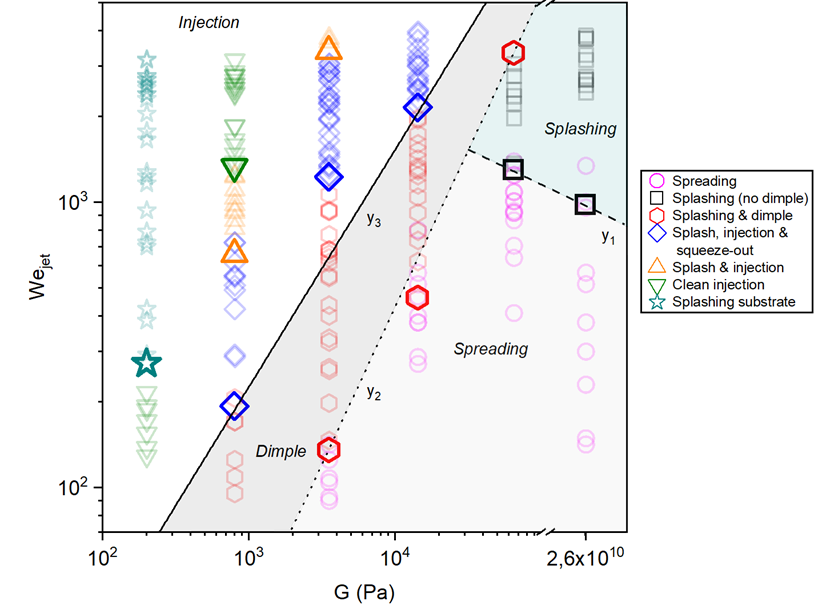}
\caption{\label{fig9:wide} Blurred symbols show all experimental conditions, while larger symbols indicate the first value at which a transition to a new Regime was seen. The X-axis shows $We_{jet}$ and the Y-axis shows substrate G in log-log scale. Symbol positions correspond to the substrates 0.125$\%$ agarose (G =  0.2 kPa), 0.25$\%$ agarose (G = 0.8 kPa), 0.5$\%$ agarose (G = 3.5 kPa), 1$\%$ agarose (G = 14 kPa), 2$\%$ agarose (G = 65 kPa) and borosilicate glass (G = 26 GPa). The dashed line (y1 =  1645X-0.022) shows the spreading/splashing threshold*, the dotted line (y2 = 0.017X1.1) shows the  dimple formation threshold, and the solid line (y3 = 0.728X0.34) shows the elastic/plastic deformation threshold. Please note the axis break. *y1 is not the actual plotted formulae, but a line drawn to represent the threshold.}
\end{figure*}

Applying y1 on the various substrates, indicates an increased splashing threshold for lower substrate stiffness, which was expected and reported by Howland et al\cite{ref37}. Furthermore, y2 allows calculating the required $We_{jet}$ for dimple formation in substrates with known G. Doing so for G = 0.2 kPa, we find: $We_{jet}$ = 6, indicating that all the generated jets within our experimental conditions would cause dimple formation on such substrates. Finally, y3 allows estimating the required $We_{jet}$ to cause plastic deformation in materials with varying G. For a 40 $\mu$m jet diameter to create plastic deformation and inject into substrates of G = 65 kPa (2$\%$ agarose), $V_{jet}$ should $>$ 92 m/s. Furthermore, to create elastic deformation for G = 0.2 kPa (0.125$\%$ agarose) this corresponds to $We_{jet}<$ 60, which would give $V_{jet}$ = 8.3 m/s for $D_{jet}$~=~40~$\mu$m. Unfortunately, verifying this experimentally would involve exploring new chip designs, able of generating lower velocity jets, which is out of scope for this paper.

Baxter et al., studied jet delivery efficiency in skin samples with different stiffness, which can be used to compare our findings\cite{ref29}. For $We_{jet}$ = 53000 and G~=~500~kPa they report 6$\%$ delivery efficiency, indicating a high rate of elastic material response and little deformation, i.e. close to the threshold of elastic/plastic deformation. Applying y3 for G = 0.5 MPa, gives $We_{jet}$ = 39100 as a threshold value for plastic deformation, which is $\approx$ 25$\%$ lower than 53000. Furthermore, for $We_{jet}$ = 53000 the storage modulus at the threshold is G = 0.72 MPa, which in the same order of magnitude of the stiffest skin sample they injected (G = 0.96 MPa). Thus, our model matches Baxter et al. observations on the skin, validating our findings for a larger range of materials.
 
\subsubsection*{Injection depth dependency}
The Regimes we introduced in the previous section can assist in performing microfluidic jet injections where splash-back and partial ejection of the fluid is undesired. However, to further optimize our system, a reliable dependence of jet characteristics on injection depth ( $d_{inject}$ ) and the dispersion in skin is needed. In Figure 10 we show $d_{inject}$ in terms of $We_{jet}$ (top) and $V_{jet}$ (bottom). We note that $V_{jet}$ is a better predictor for $d_{inject}$ compared to $We_{jet}$, as it shows a higher correlation to a linear fit, especially for 1$\%$ agarose (R² = 0.8819 versus R² = 0.1015) and 0.5$\%$ agarose (R² = 0.8349 versus R² = 0.6453). A linear relationship between $V_{jet}$ and $d_{inject}$ was previously reported\cite{ref13,ref28,ref29}. In contrast, for $We_{jet}$ and $d_{inject}$, there is no reported linear relationship. 
The poor correlation coefficient of $d_{inject}$ with $We_{jet}$ is because $We_{jet}$ depends on both $D_{jet}$ and $V_{jet}$. Therefore, the same $We_{jet}$ can result in jets with different $V_{jet}$. For example, a jet with $D_{jet}$ four times larger than another jet, their velocity would be different by a factor of two. 

To inject in substrates with G $>$ 3.5 kPa, jets with high $We_{jet}$ ($>$ 2500), caused by high $V_{jet}$ and low $D_{jet}$, are required to deform the surface. Such jets have a higher impact pressure due to their low surface area compared to high $We_{jet}$, resulting from average $V_{jet}$, yet high $D_{jet}$. Our observations align with previous studies of liquid jets impacting on liquid pools, where $D_{jet}$ did not influence the final cavity depth, but the jet momentum did\cite{ref68}. Therefore, we conclude that $V_{jet}$ better predicts $d_{inject}$ as compared to $We_{jet}$.  

Figure 11 shows the evolution of the injection depth in time for different agarose concentrations. Prior to impact, $V_{jet}$~ 58 ± 11 m/s for all the experiments. The jet decelerates after impact onto the agarose gel surface. This deceleration increases with the concentration of agarose. For example, in agarose 2$\%$ the jet velocity in its traveling direction is~0~m/s as soon as it contacts the gel at t ~ 60 $\mu$s. In contrast, for agarose 0.125$\%$, the velocity of the jet in the cavity is ~1/2 $V_{jet}$ remains almost constant  until t $\sim$ 400 $\mu$s. 
\begin{figure}[ht!]
\includegraphics[width=\columnwidth,height=\textheight,keepaspectratio]{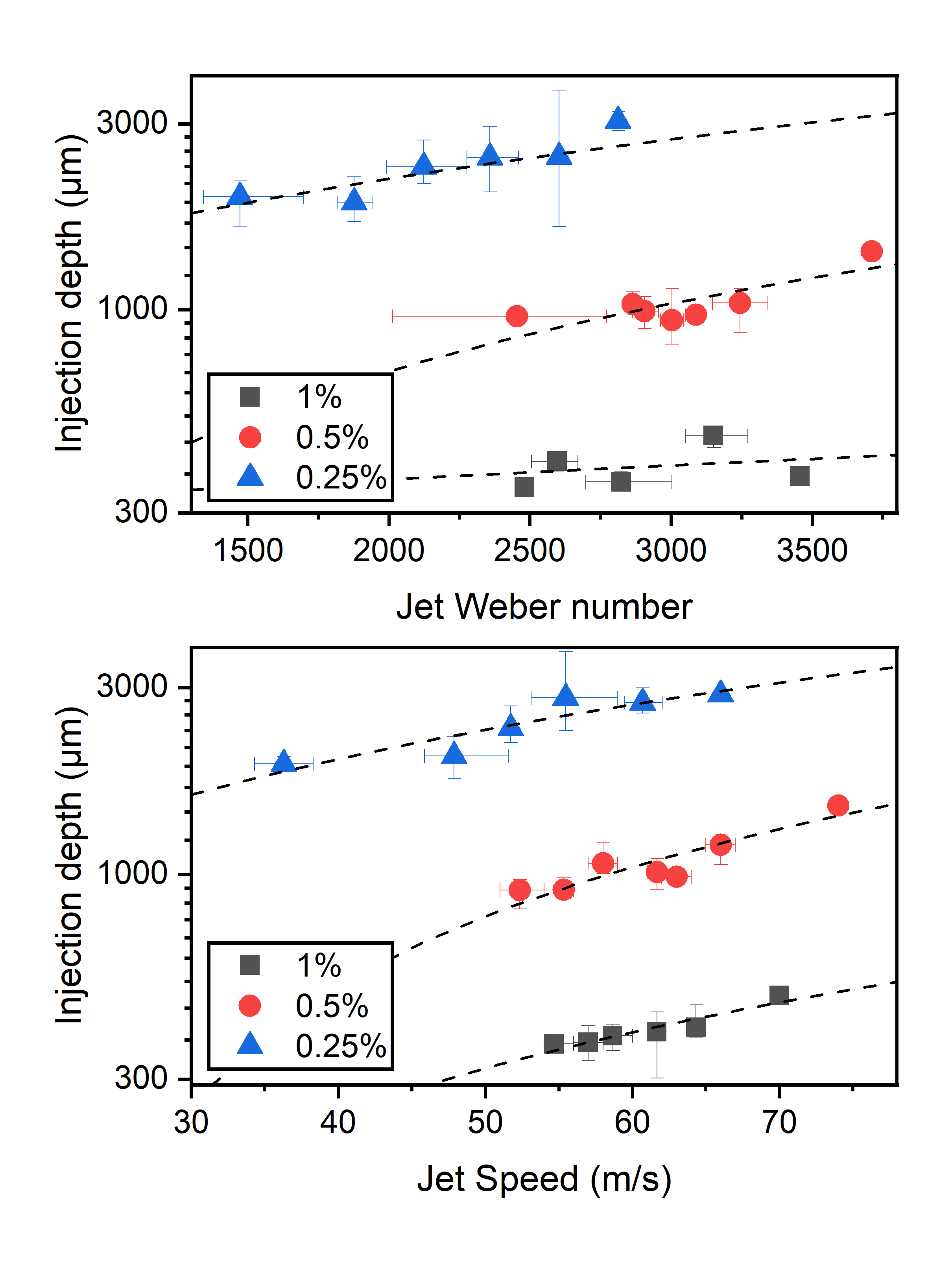}
\caption{\label{fig10} Top) Relation between $We_{jet}$ and $d_{inject}$ for 1\% agarose (G = 14 KPa), black; 0.5\% agarose, red (G = 3.5 KPa ) and 0.25\% agarose (G = 0.8 KPa). Bottom) Relation between $V_{jet}$ and $d_{inject}$ for 1\% agarose, black; 0.5\% agarose, red and 0.25\% agarose, blue. Where possible, jets within 10\% accuracy regarding $We_{jet}$ and $V_{jet}$  were grouped (n = 3). X-error bars depict the minimum and maximum values for either $We_{jet}$ or $V_{jet}$, whereas the Y-error bars depict the minimum and maximum injection depth found for the grouped jets. The dashed lines show the linear fit.}
\end{figure}

\begin{figure}[ht!]
\includegraphics[width=\columnwidth,height=\textheight,keepaspectratio]{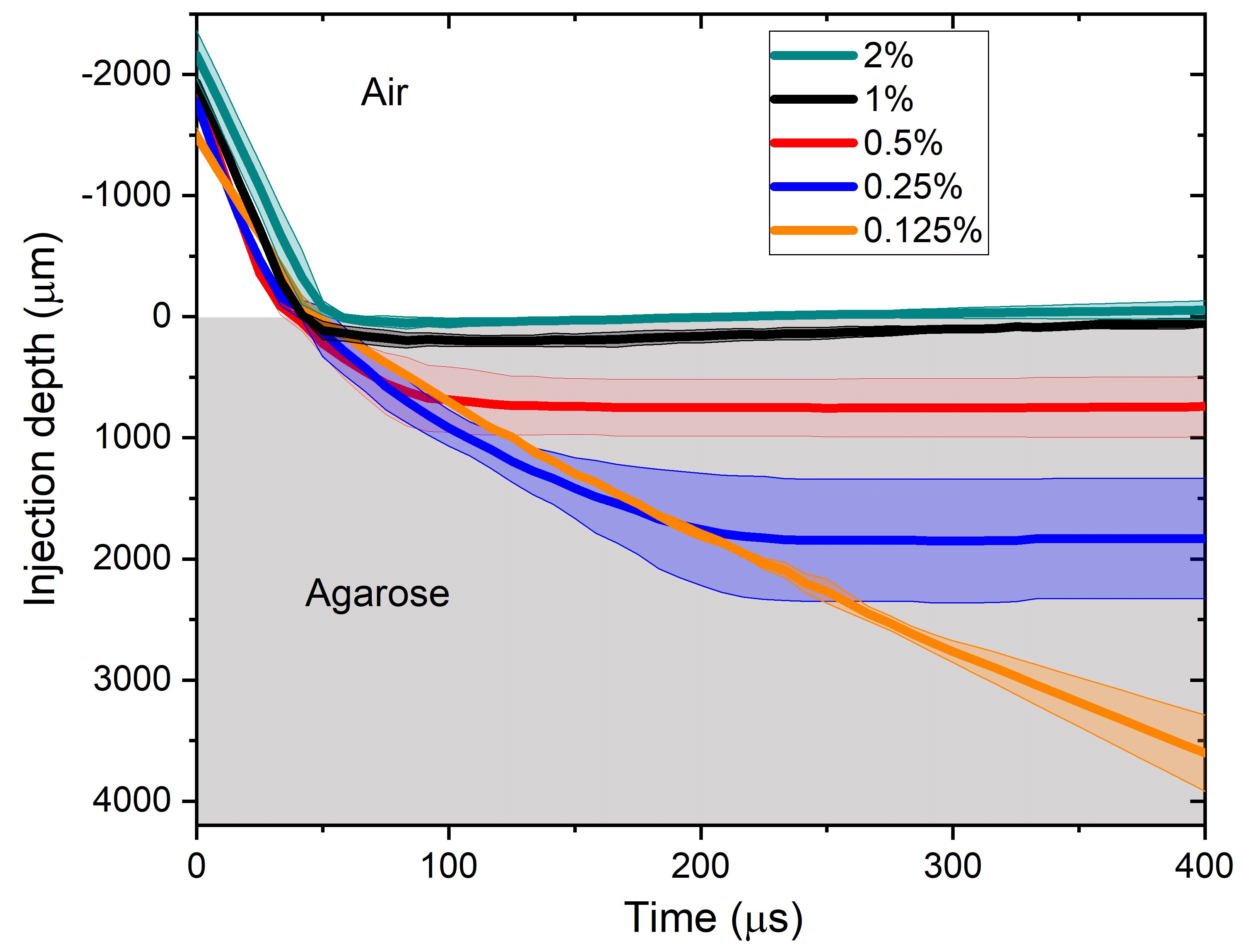}
\caption{\label{fig11} Evolution of the injection depth in time for the different agarose gel concentrations ($V_{jet}$ ~ 58 ±11 m/s, $We_{jet}$ = 3142 ± 20$\%$). The deceleration of the jet is greater for increasing agarose concentration. The solid lines indicate the average of 3 experiments, and the shaded area represents the variation across the experiments.}
\end{figure}

\section{\label{sec:level4}Conclusions}
We studied the impact of liquid microjets with a wide range of diameters (35 - 120 $\mu$m) and speeds (7 - 70 m/s), on substrates with a shear modulus between 0.2~kPa~-~26~GPa. Seven distinct regimes were identified based on the jet Weber number (70 - 4000) and substrate stiffness: 1) Jet spreading, 2) Jet splashing, 3) Surface defeat, 4) Splash, injection and squeeze-out, 5) Splashing and injection, 6) Clean injection without splashing, and 7) splashing substrate (no jet splashing).

These findings allowed us to calculate the thresholds between spreading and splashing, the threshold for dimple formation and the threshold between elastic and plastic material deformation. These thresholds can be used to predict the Weber number necessary to make a jet injection into a substrate of a known storage modulus. This result is especially relevant for needle-free injections as skin storage modulus differs between people or regions of the body. Additionally, understanding how jet Weber number relates to splashing and squeeze-out can help optimize liquid jet injections by reducing both phenomena.

We found that jet velocity is a better predictor for the injection depth compared to the Weber number, especially for the higher agarose percentages, as the Weber number depends on both jet velocity and diameter. To deform substrates with G $>$ 3.5 kPa, high impact pressure is required that is only provided by jets with high velocity and small diameter.

Although agarose gels cover the same shear modulus range as skin, they lack the complex elements of skin that cause the complex mechanical behaviour, and therefore it should be verified how the witnessed phenomenon in agarose and glass relate to skin. Future work could also be aimed at increasing the range of jet velocities to expand the regime map, and verify to what extent our calculated thresholds have predictive value over skin samples with varying shear moduli. 

\section*{Competing interest}
D.F.R. is co-founder of FlowBeams at the University of Twente, a spin-off company working on needle-free injections. D.M. is Nanovea regional Europe manager. Nanovea is a company that designs and manufactures instruments for materials testing. 

\section*{Declaration of Competing interest}
The authors declare that they have no known competing financial interests or personal relationships that could have appeared to influence the work reported in this paper.

\begin{acknowledgments}
The authors acknowledge the funding from the European Research Council (ERC) under the European Union’s Horizon 2020 Research and Innovation Programme (Grant Agreement No. 851630), and NWO Take-off phase 1 program funded by the Ministry of Education, Culture and Science of the Government of the Netherlands (No. 18844). The authors would like to thank J.J. Schoppink for both the contact angle measurements and discussions. The authors are thankful for the insightful discussions with K. Mohan and S. Schlautmann. 
\end{acknowledgments}

\appendix

\section{Agarose Young's modulus determination}
The Youngs’ modulus of the lower agarose concentrations (0.125 and 0.25\%) was calculated using the trend of the first tree samples, as depicted in Figure 12. For 0.5, 1 and 2\% the stiffness was 195.09 ± 23.29, 43.31 ± 1.42 and 10.57 ± 0.79 KPa, measured using indentations as described in the experimental section. The relation between stiffness and agarose concentration was determined to be: $y=44.352x^{2.103} (R^{2}=0.9997)$ using an exponential fit, which was used to calculate the stiffness of 0.125 and 0.25\%, being 0.6 and 2.4 KPa respectively.
\begin{figure}[h]
\includegraphics[width=\columnwidth,height=\textheight,keepaspectratio]{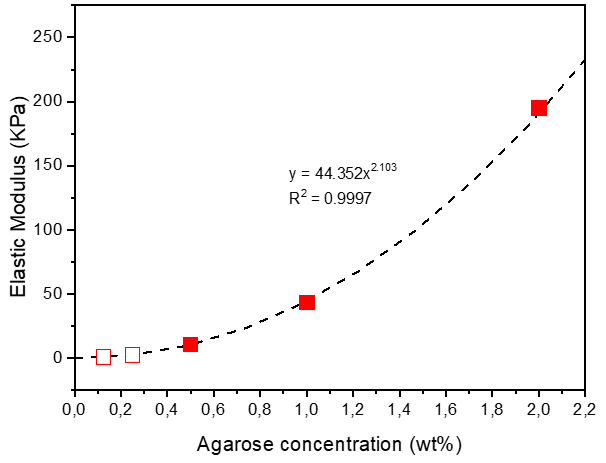}
\caption{\label{figS1} Agarose elastic modulus (Young’s modulus) in KPa as function of agarose concentration (wt\% in ${d}H_{2}O$). The solid squares indicate the values about with indentation studies (2\% = 195 KPa, 1\% = 43 KPa and 0.5\% = 10.6 KPa) whereas the open symbols indicate the values (0.25\% = 2.4 KPa and 0.125\% = 0.6 KPa) that where calculated using the exponential fit of measured values $y=44.352x^{2.103} (R^{2}=0.9997)$.}
\end{figure}

\input{Paper1.bbl}
\end{document}

%% file: Paper1.bbl
%